*Room-temperature silicon platform for GHz-frequency nano-electro-opto-mechanical systems*

D. Navarro-Urrios[1,2], M. F. Colombano[1], G. Arregui[1], G. Madiot[1], A. Pitanti[3], A. Griol[4], T. Makkonen[5], J. Ahopelto[5], C. M. Sotomayor-Torres[1,6] and A. Martínez[4]

[1] *Catalan Institute of Nanoscience and Nanotechnology (ICN2), CSIC and BIST, Campus UAB, Bellaterra, 08193 Barcelona, Spain*

[2] *MIND-IN2UB, Departament d'Electrònica, Facultat de Física, Universitat de Barcelona, Martí i Franquès 1, 08028 Barcelona, Spain*

[3] *NEST, Istituto Nanoscienze – CNR and Scuola Normale Superiore, Piazza San Silvestro 12, I-56127, Pisa*

[4] *Nanophotonics Technology Center, Universitat Politècnica de Valencia, Spain*

[5] *VTT Technical Research Centre of Finland Ltd, P.O. Box 1000, FI-02044 VTT, Espoo, Finland*

[6] *Catalan Institute for Research and Advances Studies ICREA, 08010 Barcelona, Spain*

*Abstract*

Nano-electro-opto-mechanical systems enable the synergistic coexistence of electrical, mechanical and optical signals on a chip to realize new functions. Most of the technology platforms proposed for the fabrication of these systems so far are not fully compatible with the mainstream CMOS technology, thus hindering the mass-scale utilization. We have developed a CMOS technology platform for nano-electro-opto-mechanical systems that includes piezoelectric interdigitated transducers for electronic driving of mechanical signals and nanocrystalline silicon nanobeams for enhanced optomechanical interaction. Room temperature operation of devices at 2 GHz and with peak sensitivity down to 2.6 cavity phonons is demonstrated. Our proof-of-principle technology platform can be integrated and interfaced with silicon photonics, electronics and MEMS devices and may enable multiple functions for coherent signal processing in the classical and quantum domains.

*Introduction:*

Recent advances in nanotechnology have enabled the generation, manipulation and detection of coherent motion in nanoscale devices using both optical and electrical signals. Such devices, usually termed nano-electro-opto-mechanical systems (NEOMS) [1], hold the promise for disruptive applications in the classical and quantum realms, including, e.g., microwave photonics [2], sensing [3], electro-optical modulation [4], and coherent microwave-to-optics interfaces [5]. The realization of on-chip NEOMS requires several features to be simultaneously fulfilled. To ensure strong optomechanical (OM) interaction, transparent materials able to confine photons and phonons in the same small volumes are needed. This usually requires released high refractive index films properly structured to form either waveguides or cavities with enough overlapping of the photonic and phononic fields [4],[6]-[7]. In such optomechanical crystals [8], cavities designed to confine light at telecom wavelength typically support mechanical modes at several GHz frequencies.

The conversion of mechanical waves from electrical signals can be achieved by means of metal-contacted piezoelectric materials. Forming interdigitated transducers (IDT) enables generation of surface acoustic waves (SAW), the frequency of which depends on the IDT period [9]. The generated SAWs can be further converted to volume mechanical vibrations and vice versa [10], and interact with optical fields confined in the cavity. IDTs with sub-micrometer periods can excite SAWs at frequencies of several GHz, which can couple to the mechanical modes in the OM crystals cavities. This microwave regime is highly relevant for classical and quantum applications, since most wireless networks deployed so far operate at GHz frequencies and superconducting qubits have frequencies in that range. Therefore, microwave-to-optics conversion (and vice versa) using NEOMS has become a hot scientific topic in recent years.

Microwave-to-optics conversion in on-chip NEOMS has been demonstrated in several technological platforms such as aluminum nitride (AlN) [11]-[12], gallium arsenide (GaAs) [13]-[16], lithium niobate (LN) [17]-[19], and gallium phosphide (GaP) [20]-[21]. However, in order to ensure coexistence and interoperation of NEOMS with mainstream electronic, photonic and MEMS devices, it is highly desirable to implement them on a silicon platform and using CMOS-compatible fabrication processes. The realization of silicon NEOMS has a major roadblock, which is that silicon is not piezoelectric. This means that other alternatives have to be explored to perform the electromechanical (EM) operations. Some approaches make use of the low-frequency electric fields between metallic strips to exert a force that leads to mechanical motion [22][23]. However, this approach is highly inefficient, requires very large electric powers to operate and cannot reach the microwave frequency regime. Increasing the efficiency by bringing the metal strips closer would result in higher optical losses. A different technique consists of the integration of piezoelectric material, such as AlN, on a silicon wafer. This approach has been recently used to demonstrate EM and OM interaction in silicon waveguides, leading to broadband non-reciprocal behavior [24]. More recently, combination of either AlN and IDTs [25] or microwave resonators [26] with released crystalline silicon nanobeams has resulted in successful demonstration of microwave-to-optics conversion at cryogenic temperatures.

In this work, we demonstrate an alternative silicon platform for NEOMS using Al IDTs on piezoelectric AlN layer to generate coherent GHz frequency mechanical waves that efficiently interact with optical fields in released OM cavities operated at room temperature. The core material to guide and confine both photons and phonons is nanocrystalline-silicon (nc-Si), which has shown optical and mechanical properties similar to those of crystalline silicon (c-Si) [27]. Advantages over c-Si include: i) The optical, mechanical and thermal properties of nc-Si can be tuned by annealing when transforming the amorphous Si film to nanocrystalline (for details, see SI) as the properties depend on the grain size [27]-[29], ii) the thickness, stress, grain size and doping can be tuned independently, adding extra degrees of freedom in the material and device design and iii) nc-Si NEOMS are processed on standard silicon wafers, which are less expensive than silicon-on-insulator wafers (SOI) typically used in the fabrication of silicon OM circuitry. We demonstrate microwave-to-optics transduction using 2 GHz mechanical mode excited by SAW and coupled to a 200 THz optical mode in a nc-Si fishbone nanobeam [32]. Our platform paves the way towards realization of room-temperature NEOMS compatible based-on CMOS technology for a wide variety of applications.

*Description of the technology platform:*

Figure 1 shows the experimental setup for characterizing the NEOMS a scanning electron microscope (SEM) images of one of the tested devices (top right inset) and a zoom of the OM cavity and the waveguide for evanescent light coupling (bottom right inset). The design of the OM crystal is based on the corrugated unit cell described elsewhere that displays a full phononic bandgap around 4 GHz and several additional phononic pseudo-gaps in the GHz region [30], as experimentally demonstrated in Refs. [32],[33]. The cavity region is constructed with 12 central cells in which the pitch, the radius of the hole, the stubs length and the stubs width are varied in a quadratic way towards the center. The first three geometrical parameters progressively decrease while the latter increases. This OM cavity, in contrast to one in which the stubs width is constant, allows bringing down in energy the phononic breathing modes so that those in the 2.0-2.5 GHz region coming from the $\Gamma$ point of the second even-even band are placed in a pseudo-gap for modes of similar spatial parity [30]. This leads to two improved features in comparison with the standard design. On the one hand, the breathing modes appearing now close to 2.0 GHz, being protected by the pseudo-gap, enhance their Q-factors otherwise dominated by leakage through the mirrors. In this regard, it is also worth noticing that material losses of GHz modes in optimized nc-Si OM cavities are a factor of 2-3 lower than those made of c-Si thanks to nc-Si material tensile stress [27]-[29]. On the second hand, since the IDTs are designed to operate at 2GHz with an average bandwidth of about 0.2 GHz (see Figures S1 and S2 in the Supporting Information), they can resonantly excite those breathing modes effectively. Finally, FEM simulations of the actual cavities realized by their contour retrieved from SEM images resulted in vacuum OM coupling rates $g_0/2\pi \approx 200$ kHz, which is a large enough value to observe a wide range of OM phenomena and comparable with what is obtained in the standard design [30].

The displacement readout of the mechanical cavity mode amplitude is done by coupling light into the optical cavity using an integrated waveguide that can be accessed by a tapered fiber via an adiabatic coupler. It consists of a region in which the waveguide smoothly increases its width by 90 nm in a region placed before a 90º bend, so that its effective index matches that of the tapered fiber loop placed on top of the guide [34]. The bottom SEM picture in Figure 1 shows also details of the final part of the input integrated waveguide, which is terminated with a photonic crystal mirror based on a unit cell with an optical band-gap for TE modes similar to that of the outer cells of the OM cavity. The waveguide mirror region contains a tapered region of holes of increasing diameter that reduces the losses associated with the optical mode mismatch between the propagating and the mirror regions. These characteristics, together with an optimized relative positioning of the waveguide and mirror region with respect to the

OM cavity, enables the coupling of about 50% of the waveguide input light into the optical modes of the OM cavity [35]. The top left inset to Figure 1 displays characteristic transmission (black trace) and reflection (red trace) spectra of one of the test devices. The coupling spectral bandwidth of the waveguide adiabatic coupler is about 50 nm and its central wavelength for maximum coupling depends on the specific position the tapered fiber contacts the tapered region of the waveguide, thus enabling tunability over the whole spectral range of the tunable laser. The transmission spectrum also shows an oscillatory behavior associated to optical modes of the fiber loop and a minimum at around 1560, where all the input light has effectively being coupled into the waveguide. Indeed, the contact position is chosen to maximize the optical fiber-to-waveguide coupling in the spectral range where the OM cavity modes appear. The latter become apparent in the reflection spectrum as resonant dips in the signal reflected by the waveguide mirror. In the upper left inset, two OM cavity resonances appear around 1560 and 1590 nm with Q-factors $\approx 10^4$. It is worth noticing that the transmitted signal does not contain any information from the OM cavity, which is only embedded in the reflected signal.

The IDTs were circular to electrically excite and focus the SAW, which was converted into guided mechanical waves at the entrance of the released nanobeam. Simulation results suggest that conversion efficiencies close to 1% at 2 GHz are achievable using realistic parameters [10]. Experimental laser Doppler vibrometry results unsing 1 GHz IDTs confirmed that a SAW is generated by the IDTs and is focused into a released nanobeam, where high-intensities of mechanical guided waves are observed [36]. This implies that such mechanical waves should strongly interact with optical waves if an OM cavity is inserted in the mechanical path and if both mechanical and optical waves resonate in the cavity.

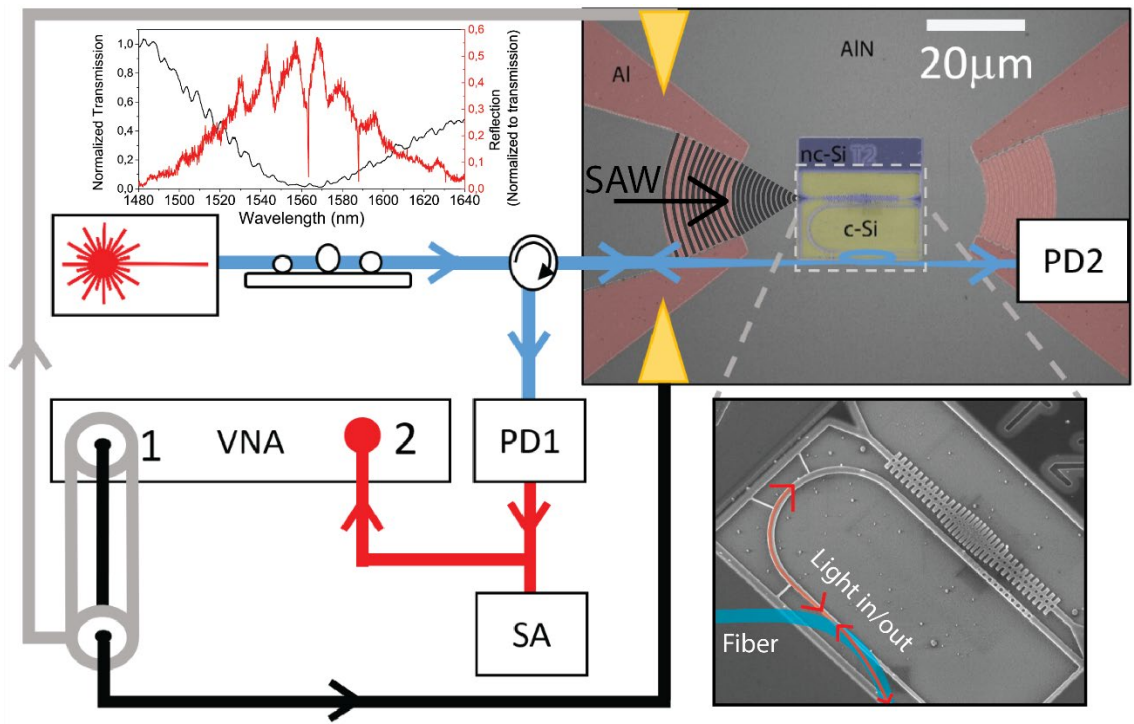

Figure 1. Experimental setup and NEOMS platform. A tunable fiber is linearly polarized and connected to a tapered fiber loop that is brought into contact with an adiabatic integrated waveguide, enabling the optical excitation of an optical resonance of an OM cavity. Reflection and transmission can be collected with near-IR photodiodes PD1 and PD2, respectively. The laser power is set to be on the order of the hundreds of μW so that radiation pressure forces can be neglected and the optical signal is only used for

*transducing the mechanical motion. The RF spectrum of the reflected signal can be extracted with a Spectrum Analyzer (SA). Coherent piezo-electrical excitation is performed with of a Vector Network Analyzer (VNA), which is connected to the input Interdigitated Transducers (IDT) through port 1. In this case, the reflected signal is connected to port 2 of the VNA, so that $S_{21}$ provides the coherent response of an OM cavity optical resonance to the piezo-electrical excitation. The top SEM image shows a device consisting of two focusing Al IDTs (with contact pads connected to metallic characterization tips) on the AlN layer (left and right parts of the image) and a released nc-Si OM circuit between them. The OM cavity is also connected to the input and output IDTs for electrical injection of coherent phonons. The bottom SEM image shows a zoom of the OM cavity coupled to a waveguide with a mirror. Details of the fiber-waveguide coupling are also shown. The top left inset shows typical transmission (black) and reflection (red) optical spectra of one of the test devices when coupled to the tapered waveguide.*

*Electro-opto-mechanical characterization*

Thermally-activated mechanical modes can be optically transduced and observed in the Spectrum Analyzer (SA) when the excitation laser wavelength is in resonance with an optical cavity mode and the modes have a minimum degree of OM coupling rate to overcome the experimental noise level. The radiofrequency (RF) spectrum measured with the SA is shown in the black curve of Figure 2 and displays a set of mechanical breathing modes of interest for this work appearing slightly above 2 GHz. Mechanical Q-factors - measured at room temperature - reach $4\times10^3$. This value is an order of magnitude higher than the values measured from samples fabricated using nc-Si with all the stubs identical on cavity modes of similar origin but not placed in a pseudo gap [28]. Moreover, the Q-factors are about a factor of 2 larger than the modes appearing at GHz range, placed either in pseudo or full gaps, in c-Si OM cavities, thus confirming the beneficial role of tensile stress in reducing mechanical losses [27],[32].

Before interpreting the measurements of optically transduced coherently driven motion, it is important to state that when the optical excitation is not in resonance with an OM cavity mode, $S_{21}$ gives values below -120 dB without significant spectral features (see Figure S1 in the Supporting Information). On the contrary, when the laser is in resonance with an optical mode such as the displayed in the lower part of the inset to Figure 2, the piezo-optic $S_{21}$ response (red curve of Figure 2) exhibits a rich structure of coherent peaks which, in some cases, overcome the noise level by 60 dB in a region consistent with the central operating frequency and bandwidth of the IDTs. Several peaks match the frequencies of the thermally activated mechanical modes (highlighted with vertical dashed lines), an example of which is illustrated in the upper part of the inset to Figure 2. The relative height of the different peaks does not match in the two experiments because, in the case of $S_{21}$, the spectral response is not only determined by the OM coupling rate as in the SA spectrum, but also by the spectral response of the IDT and the efficiency of the mechanical excitation of the specific OM cavity mechanical modes. In fact, we have observed substantial spectral differences in $S_{21}$ depending on which IDT is used for SAW excitation (see Figure S4 in the Supporting Information). It is also worth mentioning that the actual values of $S_{21}$ reflect just the ratio between the power output by port 1 of the VNA and the incoming electrical power provided by the photodetector to port 2, so it does not provide a quantitative measure of the energy efficiency of the electro-optomechanical process occurring in the device.

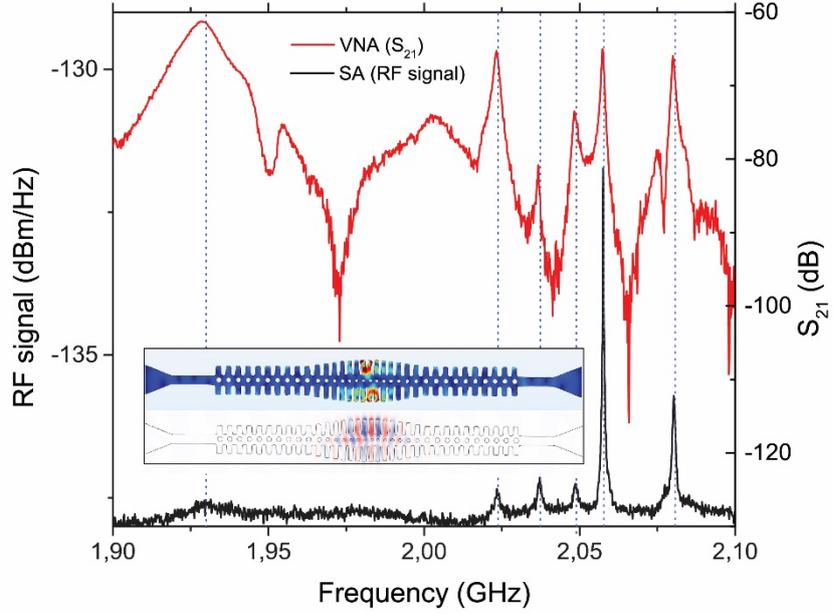

*Figure 2. Microwave-to-optics transduction from electro-opto-mechanical characterization. The black curve shows a RF spectrum measured by the Spectrum Analyzer (SA) of thermally activated modes. The red curve shows a piezo-optical $S_{21}$ coherent spectrum measured with the Vector Network Analyzer (VNA). Both curves are transduced optically by coupling light into the 1560 nm optical resonance of the OM cavity. The inset shows FEM simulations of supported optical and mechanical modes displaying OM coupling rates ($g_{OM}/2\pi$) of 200 kHz obtained by importing the geometrical profile from a SEM image of the measured device.*

*Electro-mechanical efficiency estimation*

The average phonon number of a given mode of eigenfrequency $\Omega_m$ in the high temperature (T) regime can be approximated by $\bar{n}_{th} \approx \kappa_B T / \hbar\Omega_m$, where $\kappa_B$ and $\hbar$ are the Boltzmann and Planck constants, respectively. It is then straightforward to quantify, in the SA, the average number of coherent OM cavity phonons driven by the IDT ($N_{ph}^{coh}$) by comparing the RF overall power contained in the coherent peak at the excitation frequency ($S_{coh}(\Omega_{coh})$) with that contained in the thermally activated one ($S_{th}(\Omega_m)$) for equivalent experimental conditions. The latter magnitudes are calculated by just integrating the areas beneath the RF peaks. The expression for evaluating $N_{ph}^{coh}$, which is proportionally related to the averaged squared deformation $< x_{coh}^2 >$, reads:

$$N_{ph}^{coh} = \bar{n}_{th} \frac{S_{coh}(\Omega_{coh})}{S_{th}(\Omega_m)} \quad (1)$$

In Figure 3 we estimate the electro-mechanical conversion efficiency by showing the evaluation of $N_{ph}^{coh}$ using Eq. 1 for the case of a mechanical mode at $\frac{\Omega_m}{2\pi}$ = 2.058 GHz as a function of the VNA output electrical power. The data follow a linear trend, the slope of which allows extracting an electro-mechanical conversion efficiency of 228 coherent phonons / μW, which is limited by the ohmic losses at the contacts between the metallic tips and the Al pads. This value is of the same order of magnitude reported in Ref. [13] using an IDT with an order of magnitude narrower bandwidth than in our device. Indeed, thanks to the broad IDT bandwidth, we obtain similar responses with power on the various mechanical modes that appear in the black curve of Figure 2. The linear dependence of $N_{ph}^{coh}$ with VNA output power is consistent with the observation of the invariability of the $S_{21}$ spectral response plotted in the lowest inset to Figure 3, thus implying a linear mechanical response with injected electrical power. We have also quantified the minimum number of coherent phonons that can be detected, i.e., the peak sensitivity, which is about 2.6 coherent phonons using a resolution bandwidth of 3kHz (0.05 coherent phonons/Hz$^{1/2}$) and a VNA output power slightly below 1 nW. The top inset in Fig. 3 shows this measurement, where the red and black curves display the RF spectrum around the mechanical resonance when the electrical driving is off and on, respectively. A narrow peak at $\Omega_{coh}$ in the centre of the mechanical resonance of the black curve stands out of the thermal background, thus enabling the quantification of $N_{ph}^{coh}$. Considering the electro-mechanical conversion efficiency and the sensitivity values reported above we can also compute the minimum electrical power that can be converted in a measurable optical signal, i.e. the peak sensitivity for electro-mechanical conversion, which is about 10 nW using a RBW of 3 kHz (0.2 nW/Hz$^{1/2}$). Finally, it is worth noting that, when performing the coherent piezo-optical measurement of $S_{21}$ with the VNA, the peak sensitivity improves dramatically, decreasing its value by several orders of magnitude thanks to a much lower noise level.

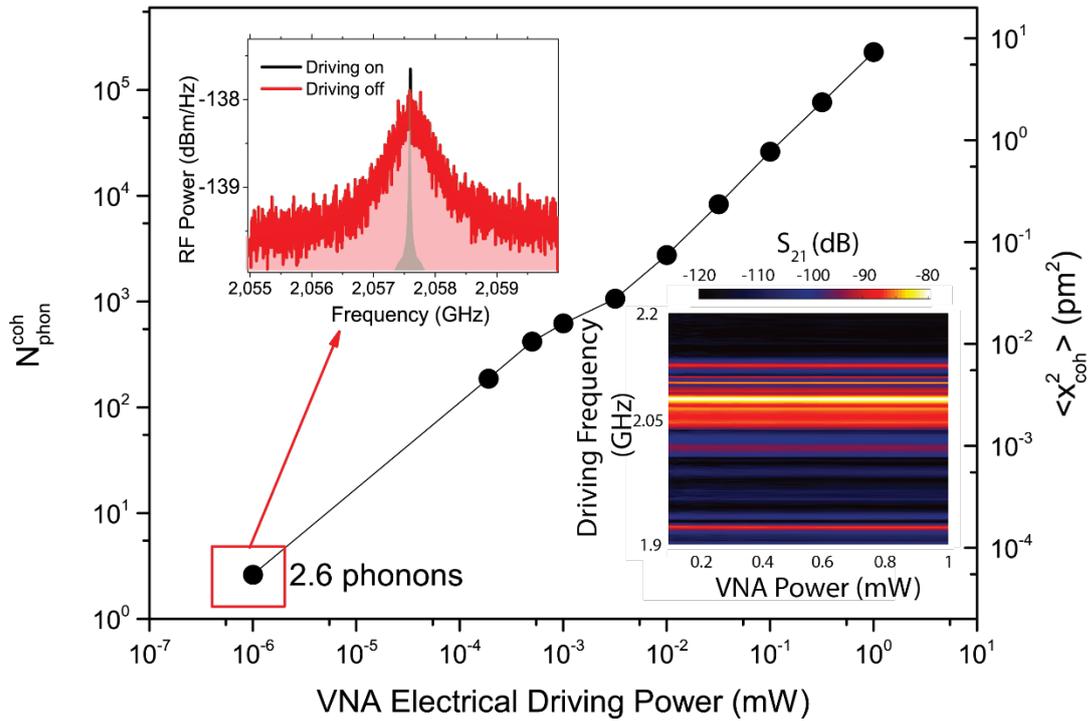

$\frac{\Omega_m}{2\pi}$ = 2.058 GHz. The right vertical axis indicates the corresponding average squared deformation. The top inset shows the RF spectra of the thermal noise when the VNA electrical driving is off (red curve) and on with a power of 1nW (black curve). The bottom inset shows the piezo-optical coherent response as a function of the VNA electrical driving power.

*Conclusions*

In summary, we have presented a new platform for on-chip NEOMS which is completely compatible with CMOS technology. Microwave-to-optical conversion at room temperature has been demonstrated with a peak sensitivity below 3 phonons. Electro-mechanical conversion efficiencies have been evaluated, leading to values of more than 200 coherent cavity phonons per microwatt of electrical power. The latter value can be greatly improved by reducing the ohmic losses in the IDT contacts, by means of wire bonding the contacts of the IDTs to the electric signal source. Our platform, which so far is a proof-of-concept, compares very well in performance to the competition [10]-[19] in terms of mechanical frequency and OM coupling rate. The key and crucial advantages are room temperature operation and complete compatibility with Si technologies, which enables coexistence with silicon electronics, photonics and MEMS. The use of nc-Si instead of its crystalline counterpart adds flexibility in tuning optical, mechanical and thermal properties of the system. The performance can be easily improved by perfectly matching the mechanical frequency of the OM cavity with the IDTs response. The frequency of the IDTs depend of the pitch of the electrode fingers and can be increased up to at least 10 GHz. At higher frequencies, series resistances and the coupling coefficient may become limiting factors. In addition to the novelty of combining AlN with Si, we have unambiguously demonstrated the possibility to excite mechanical vibrations in nanobeams using focusing of the SAWs, leading to enhanced oscillation amplitudes. The integration of AlN with a non-piezoelectric material such as silicon represents a technological breakthrough in designing and realizing NOEMS components. Furthermore, the platform developed here is based on nanocrystalline silicon, which is more affordable than crystalline silicon and more versatile, with the possibility to, e.g., fabricate multilayer systems. Besides application in quantum networks to transfer qubits via optical links, the possibility to have interacting microwave and optical signals in CMOS silicon chips operating at room temperature opens new avenues in integrated microwave photonics [2], with prospects for application in all-optical processing in wireless networks [37]. Finally, these results are very encouraging to advance the use of multi-state variables in a single chip for optimal information transmission and processing.